\documentclass[USenglish,pre,
10pt,twocolumn]{article}
\usepackage[latin1]{inputenc}
\usepackage{babel}
\usepackage{vmargin}
\usepackage
{graphicx}
\usepackage{epstopdf}

\usepackage{amsmath}
\usepackage{amsthm}
\usepackage{amssymb}
\usepackage{textcomp}

\usepackage{enumerate}

\setpapersize{A4}
\setmarginsrb{2.5cm}{0.5cm}{2.5cm}{3.5cm}{1.5cm}{1cm}{1.5cm}{1.5cm}

\newcommand{\G}{\mathcal{G}}

\newcommand{\ws}{{\wedge\star}}

\newtheorem{definition}{Definition}
\newtheorem*{definition*}{Definition}
\newtheorem{proposition}{Proposition}
\newcounter{EC}
\newtheorem*{definitionEC*}{Definition EC-\theEC}

\title{Epistemic communities: description and hierarchic categorization 
\author{Camille Roth\footnote{CREA (Center for Research in Applied
    Epistemology), CNRS/Ecole Polytechnique, 1 rue Descartes, 75005 Paris, France.
    Corresponding author: \emph{roth@poly.polytechnique.fr}} , Paul
  Bourgine\footnotemark[\value{footnote}]}}
\begin{document}
\maketitle \abstract{\small 

Social scientists have shown an increasing interest in understanding the structure of knowledge communities, and particularly the organization of ``epistemic communities'', that is groups of agents sharing common knowledge concerns.
However, most existing approaches are based only on either social relationships or semantic similarity, while there has been roughly no attempt to link social and semantic aspects. In this paper, we introduce a formal framework addressing this issue and propose a method based on Galois lattices (or concept lattices) for categorizing epistemic communities in an automated and hierarchically structured fashion. Suggesting that our process allows us to rebuild a whole community structure and taxonomy, and notably fields and subfields gathering a certain proportion of agents, we eventually apply it to empirical data to exhibit these alleged structural properties, and successfully compare our results with categories spontaneously given by domain experts. 
\newline

}

  
\section*{Introduction}

There has been recently an increasing interest from social scientists for methods of knowledge community analysis and particularly to understand their structure. To this end, several conceptual frameworks as well as automated processes have been proposed for finding groups of agents or documents related by common concepts or concerns, notably in mathematical sociology \cite{bata:gene,mood:stru,newm:dete}, scientometrics and knowledge discovery in databases (KDD) \cite{hopc:natu,lelu:extr,roch:semi}.

The focus is often on scientific communities as a large amount of data available, and in particular and among others on biologist communities --- biology is a domain where the need for such techniques is the most pressing since article production rate is currently so high that it is hard for scientists to know their community extent and to keep track of its evolution. 
In this view, it is of utmost interest to propose tools enabling agents to understand the structure and the activity of the community of knowledge they are member of, also called \emph{epistemic community}. Existing approaches in community finding are either based only on social relationships, with community extraction methods stemming from graph theory applied to social networks \cite{newm:dete,radi:defi}, or based only on semantic similarity, namely clustering methods applied to document databases where each document is considered as a vector in a semantic space \cite{lelu:extr}. 

However, there has been roughly no attempt to link social and semantic aspects, while the various characterizations of an epistemic community \cite{cowa:expl,dupo:orga,haas:intr} insist on the fact that such a community is a group of agents who share and are working on a given subset of concepts, thus suggesting that we absolutely need to take into account this duality, that is, that it is made of agents \emph{and} common interests --- agents having common interests.
In this paper, we give a formal framework for describing epistemic communities and then, we propose a method using Galois lattices \cite{barb:ordr} as well as relevant criteria for categorizing these communities in an automated and hierarchically structured fashion. Suggesting that our process allows us to rebuild a whole community structure and taxonomy, we eventually apply it to empirical data and eventually compare our results with the expected categories spontaneously given by domain experts.

Our main source of data is MedLine, a database maintained by the US National Library of Medicine and containing more than 11 million references to health sciences articles published in about 3,700 journals worldwide. Besides, we narrow our study to articles dealing with the \emph{zebrafish}, a fish whose embryo is translucid and fast developing, therefore widely used as a model animal by embryologists. 

\section{Epistemic communities}

\subsection{Rationales}

Several works stemming from social epistomology to political science and economics have given an account of the collaboration of agents within the same epistemic framework and towards a given knowledge-related goal (namely knowledge creation or validation) within what is also called an \emph{epistemic community}. For social epistemologists, it is a scientist group producing knowledge and recognizing a given set of conceptual tools and representations --- the ``paradigm'', according to Kuhn \cite{kuhn:stru} --- possibly working in a distributed manner on specialized tasks \cite{gier:scie,schm:soci}. Considering a whole knowledge field as a huge epistemic community (e.g. biology, linguistics), one can see subdisciplines as smaller embedded and more specific epistemic communities, being subfields within a paradigm.
Haas \cite{haas:intr} introduced the notion of {epistemic community} as \emph{``a network of knowledge-based experts (...) with an authoritative claim to policy-relevant knowledge within the domain of their expertise''}.
Cowan, David and Foray \cite{cowa:expl} added to this definition the fact that an epistemic community must \emph{share} a subset of concepts. In particular, an epistemic community is \emph{``a group of agents working on a 
commonly acknowledged subset of knowledge issues and who at the very least accept a commonly understood procedural authority as essential to the success of their knowledge activities''}. The ``common concern'' aspect has been emphasized by Dupouet, Cohendet and Creplet \cite{dupo:orga} who define an epistemic community as \emph{``a group of agents sharing a common goal of knowledge creation and a common framework allowing to understand this trend''}.
These authors nevertheless acknowledge the need of a notion of authority and deference. 

In the prospect of knowing which agents share the same concerns and work on the same concepts, and which these concerns or concepts are, we are farther from the epistemological point of view and need not characterize authoritative groups and their role.  
Hence, the previous definitions seem to be too precise in respect of authoritative and normative properties whereas they lack the ability to formalize accurately community boundaries and extents. Obviously such a community of knowledge should not necessarily be socially linked: it needs for instance neither be a real department nor a group of research. The definition must also allow some flexibility in the sense that an agent (or a concept) can belong to several communities.
\label{par:rationales}
We keep the idea of having common ``knowledge issues'', while we add \emph{maximality} to our definition:
\refstepcounter{EC}
\begin{definitionEC*}[Epistemic community]\label{def:ec1}
Given a set of agents $S$ 
and considering the concepts they have in common, the \emph{epistemic community} of $S$ is the \emph{largest} set of agents 
who also share these concepts. \end{definitionEC*}This conception is to be compared with the notion of \emph{structural equivalence} introduced in sociology by F. Lorrain and H. White \cite{lorr:stru} for describing a community as a group of people related in an identical manner to a set of other people -- when extending this notion to a group of people related identically to the same concept set.

Definition EC-\ref{def:ec1} is based on an agent set, and we could actually define correspondingly an epistemic community by starting from a given set of concepts, i.e. define it as the set of concepts which are at least used by the very agents that were using this given concept set. For the sake of clarity however, in the following section, we will at first focus on agent-based epistemic communities, keeping in mind that concept-based notions are defined strictly equivalently and in a dual manner (see def. \ref{def:conceptbased} below). 

\subsection{Definitions}

This being granted, we introduce from here a formal framework allowing to work on these notions. We present first the following basic definitions:

\begin{definition}[Intent]\label{def:intent}
The \emph{intent} of a set of agents $S$ is the set of concepts which are used by every agent in $S$.
\end{definition}

\begin{definition}[Epistemic group]\label{def:eg}
An \emph{epistemic group} is a set of agents provided with its intent, i.e. a group of agents and the concepts they have in common.
\end{definition}

\begin{figure}
\begin{center}
\includegraphics[width=4.5cm]{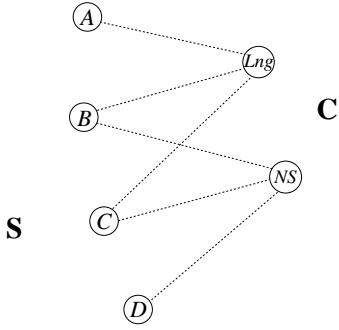}
\caption{\label{fig:reseau0}\small Sample community, and relations between agents A, B, C, D and concepts linguistics (Lng) and neuroscience (NS) (dashed lines).}
\end{center}
\end{figure}

Consider for instance that agents A, B and C work on ``linguistics'' (Lng), while ``neuroscience'' (NS) is being used by B, C and D (fig. \ref{fig:reseau0}). Therefore, the intent of \{A,B\} is \{Lng\}, that of \{B,C,D\} is \{NS\} and that of \{B,C\} is \{Lng,NS\}. Some epistemic groups of this example are thus (\{A,B\};\{Lng\}), (\{B,C\}; \{Lng,NS\}) and (\{A,D\};\{$\emptyset$\}).

If we consider a given set of agents $S$ -- notably, a group of agents  \emph{prototypic} of a field -- willing to know their epistemic community comes to identifying the greatest group of people who share the same knowledge issues as these agents (a group which thereby includes these agents).

\begin{definition}[Hierarchy, maximality]\label{def:hierarchy}
An epistemic group is \emph{greater than} another epistemic group if and only if (i) their intents are the same and (ii) the agent set of the former contains that of the latter.

An epistemic group is said \emph{maximal} if there exists no greater epistemic group.
\end{definition}

This statement allows us not only to compare epistemic groups but also and more significantly to extend a given epistemic group to its maximal social size.  
Interpreting definition EC-\ref{def:ec1} given in section \ref{par:rationales} within this framework leads now to the following definition:
\refstepcounter{EC}
\begin{definitionEC*}[Epistemic community]\label{def:ec2}
The \emph{epistemic community} based on a given agent set is the corresponding maximal epistemic group.
\end{definitionEC*}
\noindent The epistemic community based on, for instance, \{D\} is  thus (\{B,C,D\};\{NS\}), and the one based on \{A\}, \{A,B\}, or \{A,B,C\} is (\{A,B,C\};\{Lng\}).\footnote{The epistemic community based on \{B\} or \{C\} is however (\{B,C\};\{Lng,NS\}); this accounts notably for the fact that B can belong \emph{both} to a generic community and to a more specific or multidisciplinary community: (\{B\};\{Lng\}) vs. (\{B,C\};\{Lng,NS\}) -- see section \ref{sec:epistruc} for more details.}

Henceforth, with this understanding the use of relation between the set of agents and the set of concepts is sufficient to capture and describe the underlying epistemic communities of a given scientific field. By introducing an algebraic structure particularly appropriate for this purpose, Galois lattices, we offer moreover a method for representing and hierarchically grouping agents and concepts they use, which we ultimately wish to prove very relevant for epistemic community categorization.
Before doing so, we quickly introduce below the concept-based notions, defined symmetrically to the agent-based notions:
\begin{definition}[Extent, concept-based notions]\label{def:conceptbased}\label{def:extent}
The \emph{extent} of a set of concepts $C$ is the set of agents using at least every concept in $C$.
A \emph{concept-based epistemic group} is a set of concepts provided with its extent.
A concept-based epistemic group is greater than another one if and only if (i) their extent are the same and (ii) the concept set of the former contains that of the latter. A  \emph{concept-based epistemic community} is a maximal concept-based epistemic group.
\end{definition}

\subsection{Galois lattices (GL)}

Broadly speaking, using Galois lattices is possible whenever there is a relation between two sets, which are usually a set of objects and a set of properties. GL is suitable for representing and ordering abstract categories relying on such a relation, and it is therefore being widely used in conceptual knowledge systems \cite{will:con2} and formal concept classification \cite{godi:meth}.\footnote{As Wille points out \cite{will:con2}, GLs give a robust formalization of the philosophical apprehension of an abstract notion, characterized by its \emph{extent} (physical implementation) and its \emph{intent} (properties or internal content).}
In this view, considering agents as objects and concepts as properties, GL will prove to be an efficient tool to describe mathematically the notions presented above. 

Before constructing a GL we need what we call a ``pre-Galois structure''. Given two finite sets $S$ and $C$ between which we
have a binary relation ${R \subseteq S\times C}$, we introduce two 
operators ``$\wedge$'' and ``$\star$'' such that for any subset $X\subseteq S$ (resp. $Y\subseteq C$), $X^\wedge$ (resp. $Y^\star$) is the set
of elements of $C$ (resp. $S$) related through $R$ to every element of $X$ (resp. $Y$), namely:\footnote{By definition we set $(\emptyset)^\wedge=C$ and
$(\emptyset)^\star=S$.}
\begin{subequations}\begin{align}
X^\wedge&=\{\,y\in C\;|\; \forall x\in X, x R y\,\}\\
Y^\star&=\{\,x\in S\;|\; \forall y\in Y, x R y\,\}
\end{align}\end{subequations}

\begin{figure}
\begin{center}
\includegraphics[width=5.3cm]{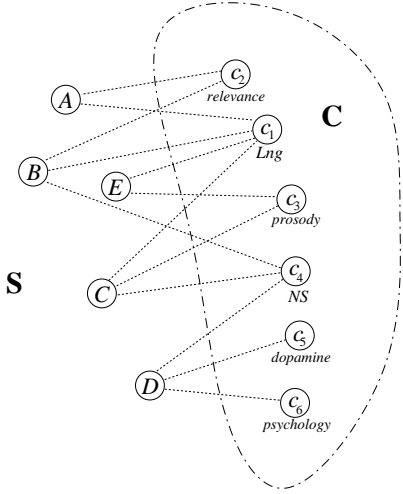}
\caption{\label{fig:reseau1} \small Extended sample community, with agents A, B, C, D and E and concepts Lng, NS, prosody (prs), relevance (rlv), imagery (img) and psychology (psy).}
\end{center}
\end{figure}

\paragraph{Interpreting preceding definitions}
Definitions \ref{def:intent}, \ref{def:eg} and \ref{def:extent} get a clear interpretation here: if $X$ is a set of agents, $X^{\wedge}$ denotes obviously its intent. Similarly if $Y$ is a concept set, $Y^{\star}$ is its extent. Thus, epistemic groups are couples of kind $(X,X^{\wedge})$ or $(Y^{\star},Y)$. 
It is also worth noting that \mbox{$X\subseteq X' \Rightarrow {X'}^\wedge\subseteq {X}^\wedge$} (expressing the fact that the intent of a bigger agent set is smaller -- the more numerous they are, the less they share) and that \mbox{$(X\cup X')^\wedge = {X}^\wedge\cap{X'}^\wedge$} (i.e. the intent of two agent sets is the intersection of their respective intents -- a group of agents has in common what its individuals share...). On the more substantial sample community described on fig. \ref{fig:reseau1}, we have for instance \{A,C\}$^\wedge$=\{Lng\} and \{NS,prs\}$^\star$=\{C\}.

Moreover, if we take the extent $X^\ws$ of an intent $X^\wedge$, that is, apply $\star$ to $\wedge$, we get \emph{all the agents} who use the same concepts that were common to the agents of $X$ (hence the largest agent set). In fact, according to definitions EC-\ref{def:ec1} and EC-\ref{def:ec2} we have:
\begin{proposition}
($X^\ws$, $X^\wedge$) is the epistemic community based on $X$.
\footnote{Indeed, (i) $X^\ws$ has the same intent as $X$ and (ii) it is the largest agent set enjoying this property. Proof: (i) comes from $((X^\wedge)^\star)^\wedge=X^\wedge$ \cite{birk:latt}; (ii) is proved by taking
$X'\supset X^{\ws}$ with $X'^\wedge=X^{\ws\wedge}$, so that $\{x\}\subset X' \Rightarrow  \{x\}^\wedge\supset X'^\wedge  \Rightarrow    \{x\}^\wedge\supset X^{\ws\wedge}    \Rightarrow   \{x\}^\ws\subset X^\ws$, but $\{x\}\subset\{x\}^\ws \Rightarrow  \{x\}\subset X^\ws$, hence $X'\subset X^\ws\qed$}
\end{proposition}

\noindent All these properties are similar and in fact dual if we consider $Y$, $\star$ and $Y^{\star\wedge}$. 

\paragraph{GL and epistemic communities} 
Besides, the operation ``$\ws$'' is a
\emph{closure operation} \cite{birk:latt}, in that it is (i) extensive (the closure is never smaller, \mbox{$X\subseteq X^\ws$}), (ii) idempotent (applying $\ws$ more than once does not change the closure, \mbox{$(X^\ws)^\ws=X^\ws$}) and (iii) increasing (the closure of a smaller set is smaller, \mbox{$X\subseteq X' \Rightarrow X^\ws\subseteq X'^\ws$}). We say that $X$ (resp. Y) is a \emph{closed} subset if $X^\ws=X$ (resp. $Y=Y^{\star\wedge}$).
Given two subsets $X\subseteq S$ and $Y\subseteq C$, a couple $(X,Y)$
is said to be \emph{closed} (or \emph{complete}) if and only if $Y=X^\wedge$ and $X=Y^\star$.  This very notion is at the core of the Galois lattice definition \cite{barb:ordr}.

\begin{definition}[GL]
  Given a relation $R$ between two finite sets $S$ and $C$, the
  \emph{Galois lattice} $\G_{S,C,R}$ is the set of every
  \emph{closed} couple $(X,Y)\subseteq S\times C$ under relation
  $R$. Thus,
  $\G_{S,C,R}=\{(X^\ws,X^\wedge)|X\subseteq S\}$.
\end{definition}

Yet such a closed couple is actually an epistemic group $(X,X^\wedge)$ where ${X^\ws=X}$. Closed couples correspond obviously to epistemic groups closed under $\ws$, and therefore it follows:

\begin{proposition}
\label{prop:epiclo}
A closed couple is an epistemic community.
\end{proposition}

This yields the fundamental property that the GL is exactly the set of epistemic communities (a graphical representation of a GL is drawn on \hbox{fig. \ref{fig:galois1}} from the sample community of \hbox{fig. \ref{fig:reseau1}}).

\section{Community categorization}

\subsection{Community structure rebuilding}
Nonetheless, if a GL contains all epistemic communities, it is still unsure whether this tool itself is meaningful or not as regards a community description task, that is, whether a GL is able to capture and reveal a given community's structure from data describing links between agents and concepts. The present section will be devoted to arguing why it can be used as such a tool. In particular, there are several stylized facts regarding the underlying community structuration we would like GLs to \emph{rebuild}, primarily the existence of subfields and significant groups of agents working within those subfields. Assuming a certain organization of scientific communities, the cornerstone of the justification of our utilization of this method will lie (i) in the fact that it does partition a field into various smaller subfields corresponding to actual scientific communities, and (ii) eventually in the agreement between epistemic communities rebuilt by GLs and those explicitly given by domain experts. 

\paragraph{Existing approaches}
Community and group detection has been for a long time under study in both computer science (graph theory as well as artificial intelligence) and sociology. Clustering methods (CM) originating from computer science tend either to use graph theory and then propose algorithms to partition graphs in a number of clusters fixed \emph{a priori} or not (such as spectral bisection or Kernighan-Lin algorithm \cite{newm:dete}), or  to consider object properties as multi-dimensional vector and endeavor to grouping objects according to their relative similarity (such as \emph{k-means} \cite{hart:kmea}, probabilistic neural networks \cite{spec:prob}, Kohonen maps \cite{koho:soms}), similarity measures being mostly euclidian distance-based. Ne\-ver\-the\-less, the main disadvantage of these methods lies in the delicate justification of their relevance for social science: they eventually produce clusters for which it is hard to tell the connection with actual sociological communities.

Approaches from sociology on the contrary introduce hypothesis and tools proper to social networks (like centrality \cite{frie:theo} or structural equivalence \cite{lorr:stru}) yielding thus CMs more adequate to social group detection than generic computer science methods, for instance hierarchical clustering \cite{john:hier}, blockmodeling \cite{bata:gene}, structural balance \cite{dore:part} or, more recently, structural cohesion and \emph{k-components} \cite{mood:stru}, and Girvan-Newman algorithm and its improvement by Radicchi \cite{radi:defi}. 

Galois lattice theory offers a convenient way to group agents with respect to concepts they share, and in this sense, it is yet another CM. Some applications of GL to social networks had also already been explored, for instance by L. Freeman and D. White \cite{free:usin} who actually apply GLs to agents and social events they attend in order to describe ``event categories''.
It is however not fortuitous to show why this very method is precisely relevant for achieving epistemic community description and categorization: in particular, for agent and concept sets large enough, a GL will contain really a lot of epistemic communities, with agents belonging to many communities with various levels of specificity.

\subsection{Epistemic community structuration}\label{sec:epistruc}

\paragraph{About relevant categorization}
Let us first examine what CMs can reveal about data: from any input set of objects provided with attributes, CMs are designed to produce an output, namely clusters of objects. However, CMs propose a grouping even when the data is a total random set of objects having almost no attribute in common, data for which any clustering would in fact be meaningless or at least irrelevant for the purpose of the study. One can try for instance sorting objects from a yard sale, e.g. according to their size and value: certainly clustering algorithms give  results, though these results are very unlikely to represent, say, functional categories. To be relevant, the use of CMs needs to be guided by particular assumptions about the data structure: a necessary assumption is obviously that it does at least exhibit a clustered structure. In other words, it is necessary to inquire and specify what a given CM aims to rebuild: it would be very imprudent to trust its output without having checked its adequacy to data and defined what really constitutes a cluster, or a community, relatively to the data. In this view, both the choice of the CM and the choice of attributes (labelling of data) are decisive.\footnote{One might thus distinguish (i) labelling irrelevant for the kind of data studied, while using a relevant CM; from (ii) CM irrelevant for the kind of data studied, however labelled relevantly. Take for instance a linguist who would like to group the words \emph{light, dark, holy} and \emph{evil} as regards their semantic field. He might consider two criteria: \emph{brightness} and \emph{goodness}, and select e.g. the following numerical representations:
light: +5 (brightness), +1 (goodness); dark: -5, -1; holy: +1, +5; evil: -1, -5.
For sure an irrelevant labelling, i.e. a bad choice in the previous criteria (say, choosing the number of vowels and the number of consonants) would obviously give him a meaningless result. But an irrelevant clustering method, e.g. based on euclidian distances, would also give him inconsistent output
in grouping light with holy, and dark with evil, while he wanted light
with dark, and holy with evil.}  

The same goes with Galois lattices: one can draw a GL from any two sets of objects and a given relation between them, but there is no reason \emph{a priori} that the lattice reveals a remarkable structure, even if it is built, represented or managed efficiently. In fact, there should exist a lot of data for which this categorization is just not relevant.
Thus, in order to know whether and why GL is an appropriate CM for producing a taxonomy of knowledge communities, it is first necessary to inquire the nature  and organization of these very communities.

\paragraph{Assumptions} Our main assumption is that there are fields of knowledge which can be described by concept lists (relevant labelling), and which are being implemented by sets of agents. Taking again the first example, some people are obviously linguists: among them, some deal with a given aspect, say prosody, while others study relevance; some other scientists deal with neuroscience, while a few of them are interdisciplinary and use both concepts. 
Knowledge fields and their corresponding agent sets are in our case  epistemic communities, which are precisely what GLs consist of (see \hbox{prop. \ref{prop:epiclo}}). Moreover and also crucial, these fields are hierarchically organized: (i) a general field can be divided into many subfields, themselves possibly having subcategories or belonging to various general fields, and (ii) some fields can be \emph{multi-disciplinary} or \emph{inter-disciplinary} in that they respectively involve or integrate two or more subfields \cite{klei:inte}. For instance, {cognitive science} is a general field gathering various subfields such as {cognitive linguistics} and {cognitive neuroscience}, thus being multidisciplinary. But the very subfield {cognitive neurolinguistics} is interdisciplinary in that it mixes and coordinates the approaches from both parent disciplines. 

\begin{figure}
\begin{center}
\includegraphics[width=8.1cm, trim=10 0 0 0]{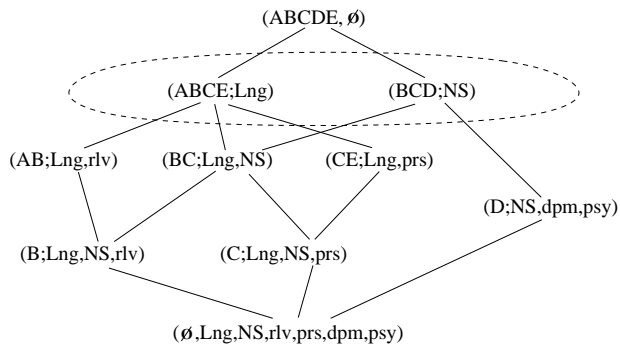}
\caption{\label{fig:galois1}\small Galois lattice of the extended sample community (hierarchical structure drawn in solid lines relatively to $\sqsubset$, i.e. ``bottom''$\sqsubset$``top''). The medium level (dashed ellipse) contains closed couples (\{A,B,C,E\};\{Lng\}) and (\{B,C,D\};\{NS\}) obviously corresponding to major fields (linguistics \& neuroscience). Hierarchy yields just below interesting subcommunities like (\{D\};\{NS,img,psy\}) or (\{B,C\};\{Lng,NS\}), possibly prototypical of more specific subfields.}
\end{center}
\end{figure}

\begin{figure}
\begin{center}
\includegraphics[width=5.75cm]{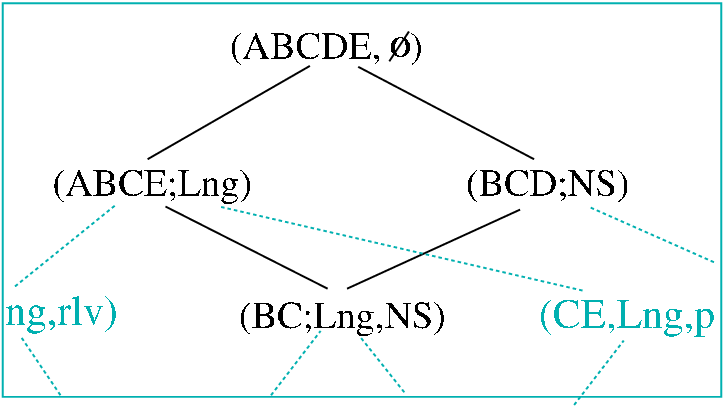}
\caption{\label{fig:diamond}\small Zoom on \hbox{fig. \ref{fig:galois1}} showing one possible \emph{\hbox{diamond}}. A multidisciplinary field is at the diamond's top (here ``$\emptyset$'', which relatively to the context can be considered as ``cognitive science'') and covers the two intermediate subfields (cognitive linguistics and cognitive neuroscience), which themselves, when combined, define an interdisciplinary subfield (cognitive neurolinguistics). }
\end{center}
\end{figure}

GL acute relevance as regards these properties results actually from its natural partial order $\sqsubseteq$ defined such that given two epistemic communities (or closed couples) $c=(X,X^{\wedge})$ and $c'=(X',{X'}^\wedge)$, we have $c\sqsubseteq c' \Leftrightarrow X\subseteq X'$. This partial order indeed makes $\G_{S,C,R}$ be a lattice, hence enjoying a hierarchical structure.\footnote{A lattice is a partially-ordered set such that any subset has a least upper bound and a greatest lower bound -- obviously a finite partially-ordered set is a lattice.
Note that the hierarchy here has nothing to do with the one introduced in def. \ref{def:hierarchy}.} More precisely, the order reflects a generalization/specialization relation, in the sense that $c\sqsubseteq c'$ means that $c$ has a smaller extent and a greater intent than $c'$, $c$ represents a smaller community dealing with a bigger concept set than $c'$, $c$ being thus more specific. This hierarchy describes exactly relations between fields and subfields as discussed in the previous paragraph (fig. \ref{fig:galois1}), as well as multidisciplinarity and interdisciplinarity through particular patterns called \emph{diamonds} (fig. \ref{fig:diamond}).

\subsection{GL and categorization}\label{sec:glcateg}
Given their hierarchical structure, GLs are thus a relevant method to list and order epistemic communities and subcommunities. However, it is still unclear why a GL, which is an ordered although possibly huge set of epistemic communities, will produce an useful and usable categorization of the community under study. A GL contains indeed all epistemic communities, a property already restrictive since agent or concept sets whose intent or extent is $\emptyset$ (i.e. they have nothing or nobody in common), or more generally is not ``closed'', are no epistemic communities and hence do not appear in GL. However, many real epistemic communities are still of no interest -- in that they do not correspond to an existing or relevant field of knowledge -- because for instance they are too small and/or too specific. In particular, for a single scientist $\{s\}$, the closure $\{s\}^\ws$ will admittedly be equal to $\{s\}$, since there are strong chances that no other scientist uses at least the same concepts as $s$ -- to some extent $s$ is ``original''. Certainly knowing that $(\{s\},\{s\}^\wedge)$ is an epistemic community is not very enlightening. If however we consider that $s$ is working on a field $F$ (i.e. $F\subset \{s\}^\wedge$), when adding more and more agents working on $F$ to $\{s\}$, as the cardinal of this agent set $S$ increases there are more and more chances that its (decreasing) extent $S^\wedge$ reaches the actual knowledge field $F$. The intent $S^\ws$ will be at this point the whole community working on $F$: there will thus be a gap between the small uninteresting epistemic communities reached hitherto, and the suddenly emerging epistemic community ($S^\ws,S^\wedge=F$). In other words, we conjecture that there is a relevant level for which closed sets $S^\ws$, and identically $C^{\star\wedge}$, are representative of a field or a trend. This also means that some epistemic communities listed by GLs are deemed to be prototypic of these fields. They are located between the whole agent set (obviously too general) and too specific communities, that is, at a medium-level of generality which is to be compared to Rosch's basic-level of categorization \cite{rosc:cogn}.

Given these assumptions, $\G_{S,C,R}$ is expected to exhibit significant structural properties -- as regards e.g. highly-populated communities, for there will be aggregate of agents around some precise fields (i.e. epistemic communities with high-size agent set will prevail). These properties, once identified, could help design criteria for detecting in a somewhat automated manner major trends (basic-level categories) within a more general field, therefore making GL a powerful categorization tool. This idea had been introduced by the present authors in a previous paper \cite{roth:bind}, now we will bring in section \ref{sec:empirical} empirical evidence to support this conjecture.

\paragraph{Comparison with existing approaches} 

In general, existing studies like those mentionned at the beginning of this section attempt to infer communities from a very general point of view (in that there is no particular assumption on the nature of the social groups that these CMs are supposed to extract from data), and still focus and rely only on single networks of social relationships (e.g. coauthorship) that may prove to be insufficient and inefficient in order to find epistemic communities which, as we said before, are not necessarily socially linked. Data duality brought by the reciprocal linkage of agents to concepts and the corresponding symmetry between agent-based and concept-based notions (\hbox{def. 1}, 2, 3 and EC-2 vs. \hbox{def. 4}) is moreover particularly well rendered by a GL, being a hierarchy of closed couples considered indifferently as agent sets or as concept sets.

It is also worth noting that some of these methods produce hierarchically structured clusters (e.g. hierarchical clustering and structural cohesion) which  seem to be close to GL hierarchical representation are in fact more or less \emph{dendrograms}. Yet, a dendrogram is a tree whereas a GL is a lattice, i.e. a generalization of trees where ascendancies can be multiple: a community is not bound to be embedded into a lineage of increasing communities, it can have ascendancies in various ``directions''; in other words, an agent can be part of many non-embedded communities, he can be to some extent ``pluridisciplinary''.

GLs are hence a particularly adapted CM for the very prospect of building knowledge community taxonomy. Moreover, although GLs are within this paper principally applied to scientific communities, we could yet easily apply it to other spheres like for instance economic communities, where companies deal with sets of technologies.

\section{Empirical results}\label{sec:empirical}

\subsection{Experimental protocol}
To lead our experiments on scientific communities, we need data stipulating which agents use which concepts. We consider article collections, assuming that articles are a faithful account what their authors are working on. 
However, an important point is now to define precisely what a
\emph{concept} is, and in particular what is a
 concept such that we can observe its appearance
in an article. This notion needs not be too precise nor too wide. Is it a paradigm like \emph{``universal
  gravitation''} or a simple word like \emph{``operon''}?   For instance,
authors provide their articles with keywords: apparently, considering
these keywords as concepts seems to constitute a relevant level
of categorization while being a convenient idea.  Yet, such
keywords have not proven to be very reliable indicators of the issues
articles are dealing with, for authors often omit important keywords
or specify poorly relevant ones; depending on the database, keywords for the same article can strongly differ, requiring the additional help of an expert ontology.

\paragraph{Word groups as concepts} Getting concepts through words and nominal groups (terms) from article title, abstract or body appears to be a safer method than using keywords. At first we will thus say that \emph{each word or nominal group is a concept} even if we are still hampered by linguistic phenomena like homonymy, po\-ly\-se\-mia, synonymy \cite{jack:foun}, syllepsis \cite{jacq:mode}, and the fact that different authors might have different definitions of the same word or understand different concepts under an identical nominal group \cite{lavi:syst}. Some techniques have been proposed (see e.g. \cite{wang:usin}) and could be used to solve these problems and determine the contextual meaning of nominal groups, this is however not the purpose of the present article and we will assume here that nominal groups represent \emph{sufficiently} distinguishable and homogenous references to concepts. 
Additionally, this definition does not prevent us from observing higher-level
concepts such as theories or even paradigms, since we can easily
refer to these concepts \emph{a posteriori} by considering sets of words, like for example interpreting \emph{\{``cell'', ``DNA'',
  ``gene'', ``genetics'', ``molecular''}\} as
\emph{molecular biology}.

We will also only proceed with title and abstract words, first because complete article contents are rarely available on an exhaustive basis (that is, exhaustively available for a whole community), and second because it could imply to take into account too many very precise though irrelevant words (thus dramatically increasing set sizes while massively introducing noise).

\begin{figure}
\begin{center}
\includegraphics[width=0.97\linewidth
]{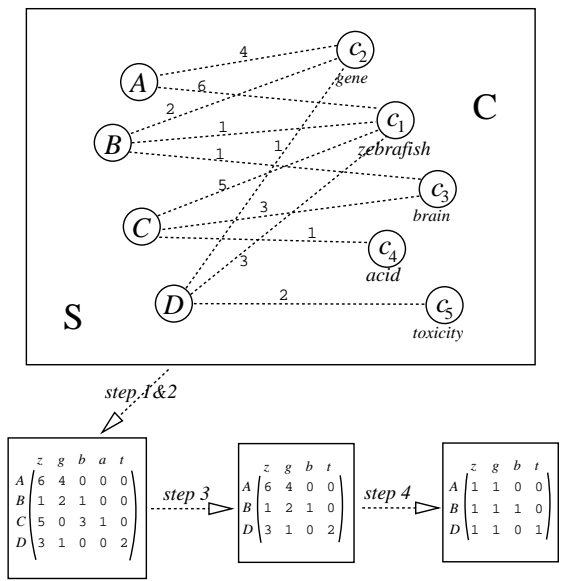}
\caption{\label{fig:protocol} \small Experimental protocol: step 1 and 2 help create the core network, and the corresponding relation weighted matrix shown here (authors on rows, concepts on columns). Some agents are removed through step 3. The GL is then computed from the binary matrix obtained after step 4. }
\end{center}
\end{figure}

\paragraph{Data processing}
The data presented here has been processed according to the following methodology: 
\begin{enumerate}
\item Collect and automatically process article data (title, abstract, authors) for a given community and period of time. As regards abstract and title, we apply a very basic linguistic processing (though a good tradeoff between complexity and efficiency) consisting in:
\begin{itemize}
	\item Excluding unsignificant words (\emph{stop-words}), such as common and rhetorical english words (``often'', ``then'', ``we'', etc.) and irrelevant words in respect of the domain (``demonstrate'', ``postulate'', ``specimen'', ``study'', etc.), using a list of more than 2,500 words, to which we add non-words such as figures, percentages, dates, etc.
	\item Excluding rare words, i.e. words appearing $n$ times or less in the whole corpus (such as words appearing only once, also called \emph{hapax legomena} or \emph{hapaxes}). In our case, we took $n=4$.
	\item Stemming the remaining words, i.e. reducing morphological variants of words to their stem (root form) using a slightly improved version of Porter's stemming algorithm \cite{port:algo}, and then creating the corresponding word classes (for example, ``genetic'' and ``genetics'' both reduce to ``genet'').
\end{itemize}
\item \label{steptwo} Identify unique authors and unique words, and then create the weighted matrix $M$ of links between authors and words, where $M_{ij}$ is equal to the number of articles where author $i$ used concept $j$ (see fig.\ref{fig:protocol}). 
\item \label{stepthree} Keep randomly a given fraction of authors, that is, consider a representative sample of the whole community by extracting randomly and uniformly some lines from matrix $M$. We chose to keep each line with probability $.25$ (this step aims only at GL reducing computation time)
\item \label{stepbin} Make $M$ a binary matrix relatively to a given threshold $\alpha$, i.e. replace $M_{ij}$ by 0 if $M_{ij}<\alpha$, else by 1: this means that an author will not be related to a concept he used less than $\alpha$ times. We actually used a threshold of 1 (increasing the threshold would critically reduce both computation costs and results significance).
\item \label{stepgal} Calculate the Galois lattice for the binary relation matrix $M$, using an implementation of Ganter's algorithm \cite{gant:algo,lind:conc}.
\end{enumerate}

\subsection{Results and comparison with random relations} 
We ran the process on articles published between 1990 and 1995 obtained through a search for ``zebrafish'' on the MedLine database, totalling 418 articles and mentioning 797 authors and 2129 words after step \ref{steptwo} of the protocol. After step \ref{stepthree}, only 218 authors and consequently 1817 concepts remained in $M$. This is the relation matrix we used for computing the GL (steps \ref{stepbin} and \ref{stepgal}). 

\begin{figure}
\includegraphics[width=1.0\linewidth, trim=65 50 0 5]{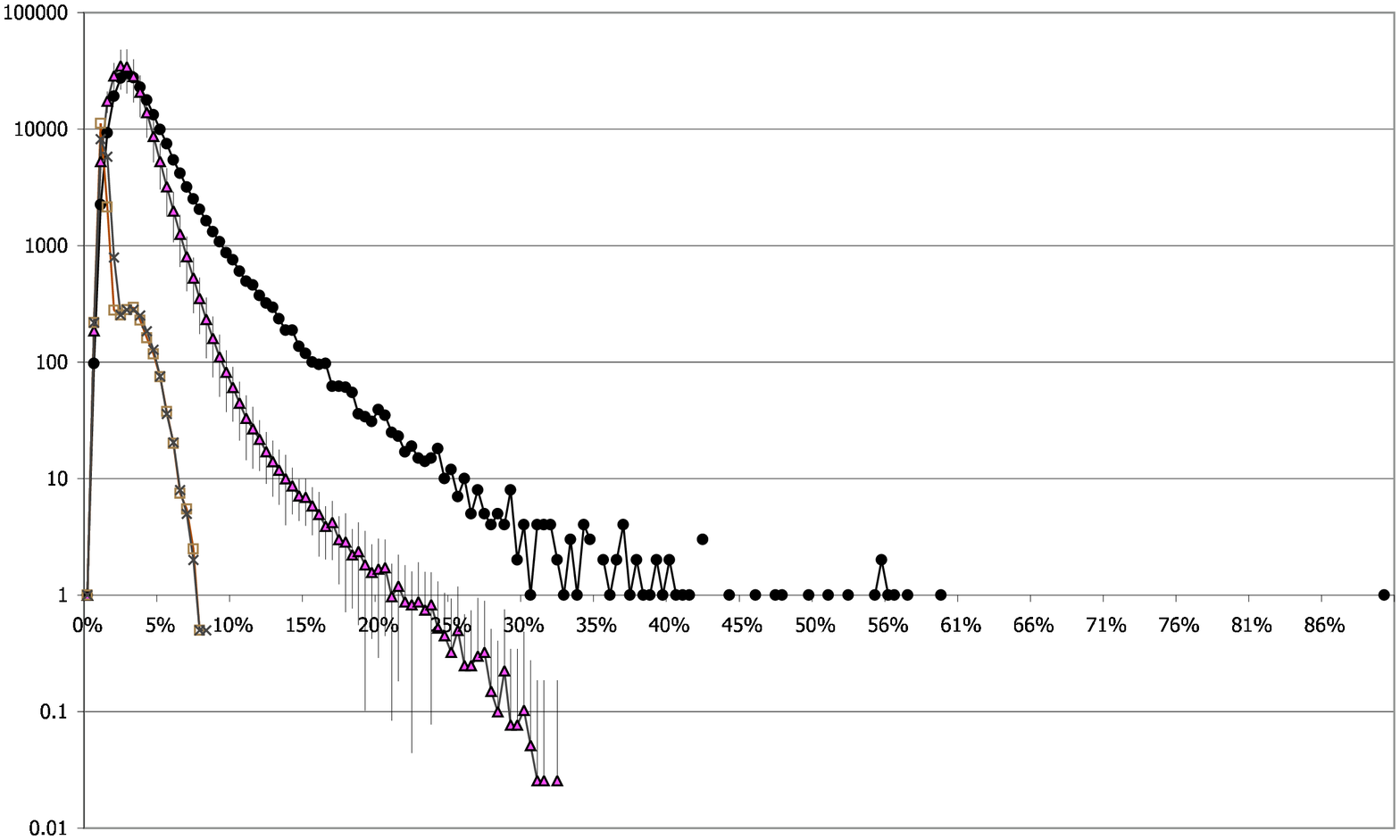}
\caption{\label{fig:distrib1} \small Raw distributions of agent set sizes (log/lin graph). Abscissa: agent set sizes (percentage of the whole community); ordinate (log scale): number of corresponding epistemic communities. Circles: empirical data; triangles: random case (random data with same distributions, 40 computations, with standard deviation bars). Also plotted on the left are two other random cases (see footnote \ref{footnote:otherrandom}): (i) random data with same link density (squares) and (ii) random data with same distribution from agents to concepts only (crosses).}
\end{figure}

We noticed unsurprisingly that some authors and concepts were appearing significantly more frequently than others. More precisely, there was a particular distribution of links from agents to concepts (proportion of agents being related to a given number of concepts) and from concepts to agents: a lot of agents (resp. concepts) were linked to few concepts (resp. agents) while few agents/concepts were related to many concepts/agents. 
For this reason, we could fear GL artefacts since frequent authors or frequent concepts are more likely to share or respectively be shared by more concepts or agents, thus being part of bigger closed sets and increasing the number of these big sets, eventually modifying artificially the GL structure, especially high-size closed sets. We hence decided to compare our results with those from GLs calculated with random-generated relations where this exact property of the empirical data was kept. In other words, we kept the distributions of links on rows and columns in the relation matrix from step \ref{stepthree} while we reshuffled the links themselves, using an algorithm introduced by Molloy \& Reed \cite{moll:rand}.\footnote{\label{footnote:random}Briefly, this algorithm consists in assigning to each author a number of outgoing links to concepts according to the desired distribution, and identically assigning to each concept a number of outgoing links to authors; then matching randomly the dangling links between authors and concepts.}
From now on, we call \emph{``random case''} the results obtained from computations on 40 such random relation matrices.\footnote{\label{footnote:otherrandom}We also considered two other random cases: (i) keep the same density in the relation (same proportion of real links in respect of possible links), which is approximately one link out of 30; and (ii) keep only the distribution of links from agents to concepts. Interestingly, the corresponding GLs are really poor: they are dramatically small, with 16,000 epistemic communities whose sizes do not exceed  5\% of the whole community in general (see \hbox{fig. \ref{fig:distrib1}}). Therefore, these cases were not investigated further.} 

\paragraph{Empirical vs. random}
In order to confirm the intuition that we have relatively large communities sharing concepts (prototypical of a subfield), we looked at the proportion of high size epistemic communities by drawing the distribution of agent set sizes.
In spite of the extremely rough linguistic assumptions, we get strongly significant results from empirical data, especially when compared to the random case. 

On the first graph (fig. \ref{fig:distrib1}) we plotted the raw distributions of agent set sizes, i.e. the number of epistemic communities relatively to the size of their agent set. The empirical GL contains 214,000 closed couples, with agent set sizes ranging from 1 to 196 -- admittedly excepting the epistemic community (S,$\emptyset$) containing all the 218 agents under study -- to be compared with an average of around 207,000 closed couples in the random case (standard deviation $\sigma\simeq64,700$), with agent set sizes ranging only from 1 to 60 ($\sigma\simeq5$). This means that while the empirical GL is generally approximately the same size as random GLs, it contains dramatically more high-size epistemic communities (featuring 371 communities representing more than a fifth of the whole agent set, when random GLs hardly contain a dozen such communities). 

\begin{figure}
\includegraphics[width=1.0\linewidth, trim=65 50 0 5]{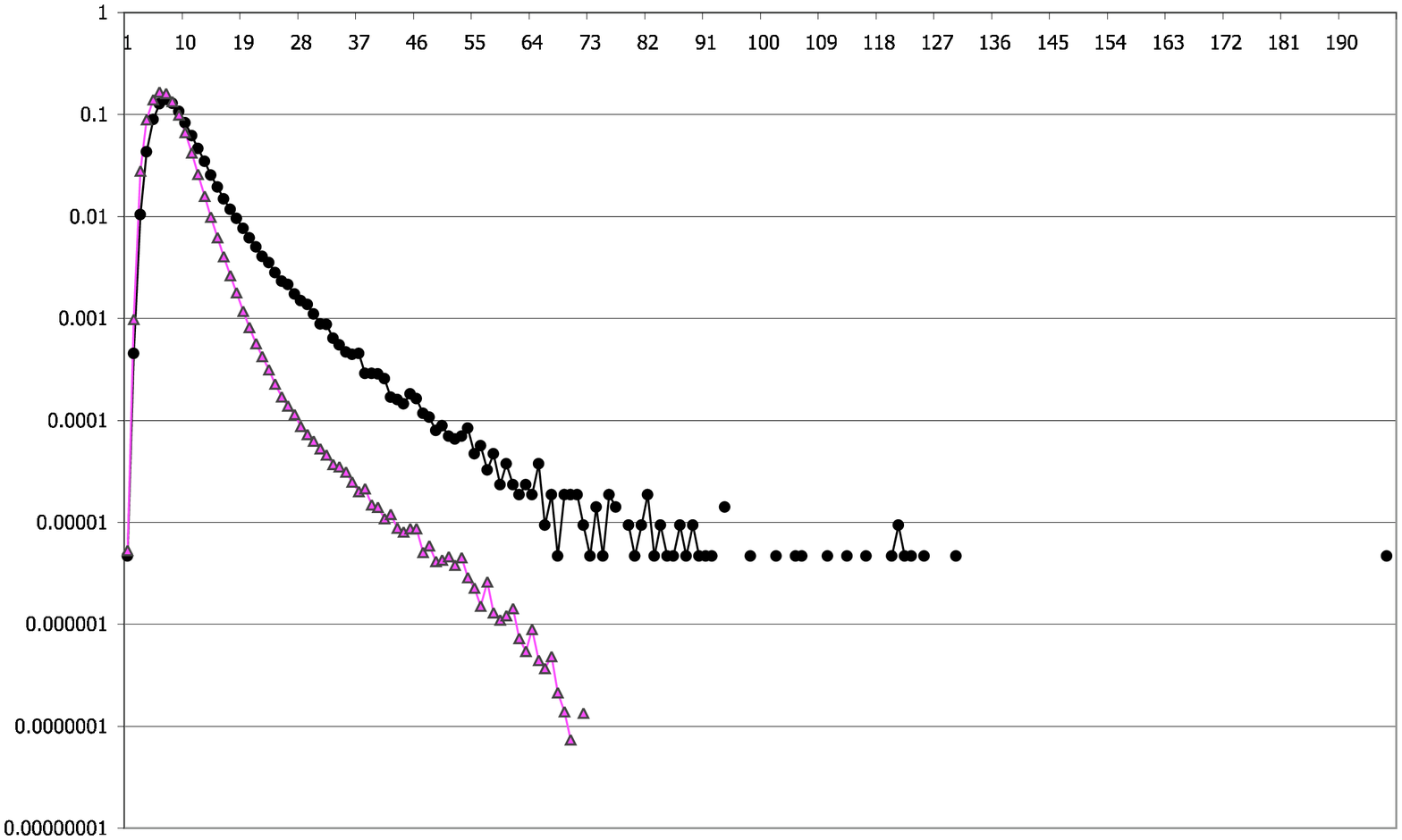}
\caption{\label{fig:distrib2} \small Normalized distributions of agent set sizes (log/lin graph). Abscissa: agent set sizes; ordinate (log scale): percentage of epistemic communities of a given size within the whole GL. Circles: empirical data; triangles: random case.}
\end{figure}

\begin{figure}
\includegraphics[width=1.0\linewidth, trim=65 50 0 5]{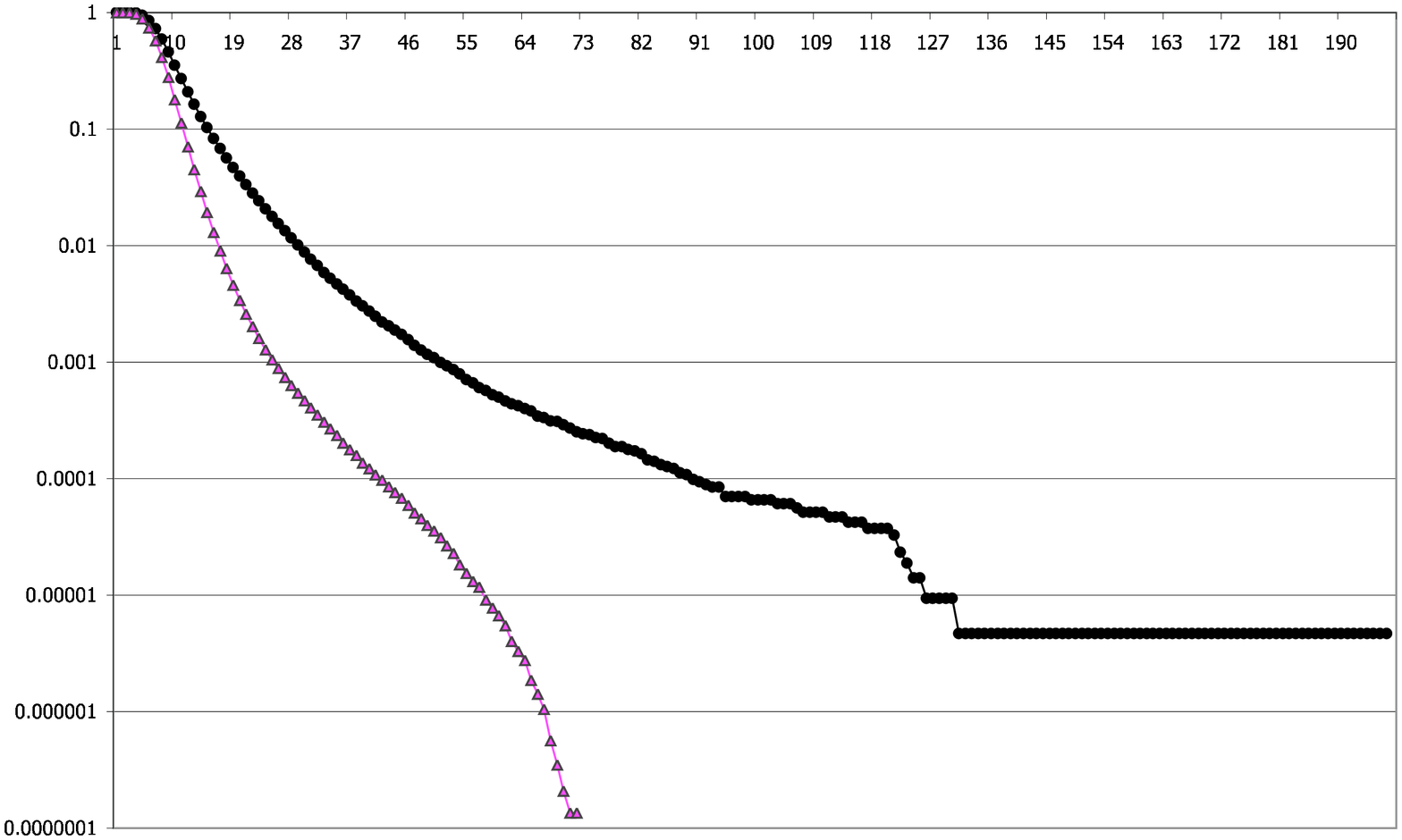}
\caption{\label{fig:distrib3} \small Cumulated densities (frequencies) of agent set sizes. Abscissa: agent set sizes; ordinate: percentage of epistemic communities containing \emph{at least} $x$ agents. Dark circles: empirical data; grey triangles: random case.}
\end{figure}

The comparison is a bit more striking on the second graph (fig. \ref{fig:distrib2}) representing distributions normalized in respect of GL size (that is, each class size has been divided by the GL total size): while there is a quite perfect fit on the density of low-size closed couples, the empirical GL is comparatively dramatically denser on high-size couples, with a deviation of one order of magnitude when considering communities with more than 20 agents, i.e. 10\% of the whole.  
For the purpose of underlining this effect, we finally considered cumulated densities on the third graph (fig. \ref{fig:distrib3}), i.e. the proportion of closed couples containing at least a given number of agents: 1\% of the GL in the empirical case is made of epistemic communities containing 30 agents or more, versus .05\% in the random case (respectively one thousandth vs. one thirty-thousandth 
for communities with 50 agents or more).


\paragraph{Rebuilding the structure}
High-size epistemic communities appear to be proper to our empirical data, suggesting that these high-size clusters --- that is, large groups of structurally equivalent agents \cite{lorr:stru} pointing to the same groups of concepts --- are a remarkable stylized fact, providing support to the conjecture outlined in section \ref{sec:glcateg}. Nonetheless, it is also of great interest to know whether these communities are significant and relevant, and notably if they help partition a field into various smaller subfields corresponding to real epistemic communities -- a stylized fact as much crucial for the justification of the utilization of this very CM.

With the help of a zebrafish expert, Nadine Peyri\'eras, we observed that it was actually the case:
\begin{enumerate}[(i)]
\item The first and biggest community is unsurprisingly centered around the word \emph{``zebrafish''} and contains 196 agents (90\% of the whole). The fact that it does not reach 100\% of the community as one would expect reflects the imperfection of the empirical data collection and processing. 
\item Then, a lot of large epistemic communities is revolving around a small set of words, namely \emph{``gene''}, \emph{``expression''}, \emph{``pattern''}, \emph{``embryo''}, \emph{``develop''} and \emph{``vertebrate''}, that is, their intents are a combination of some of these words while their extents contain generally around 100 agents. In fact, a large majority of the 218 agents are present in at least one of these communities; this word set seems accordingly to characterize the core paradigm of zebrafish researchers (even if each agent does not use it wholly, which is credible if we consider that in the relatively few article abstracts present in the database most authors might have not cited \emph{every} word of this word set but only a partial subset).  According to our expert and the litterature \cite{grun:head}, the zebrafish is indeed being used as a vertebrate animal model for the study of gene expression and function during embryonic development.

Similarly, another word subset of interest is made of \emph{``cloning''}, \emph{``stage''}, \emph{``transcription''}, \emph{``sequence''}, \emph{``protein''}, \emph{``region''}, \emph{``encode''}, which constitute the intents of relatively high-size epistemic communities (50 agents). According to our expert, these words are proper to the paradigm of molecular biology or developmental studies in general, or to zebrafish study, which consists in isolating a large number of mutant fish lines, isolating the corresponding mutated genes, then investigating their involvement in biological processes. So, in the search for relevant partitioning communities it is reasonable to ignore these too trivial thus noisy words and the corresponding closed sets.

\item Thereafter and once these words ignored, some smaller and more precise communities appear around non-paradigmatic words. Two major groups emerge first: (i) one with the epistemic community based on \emph{``growth''} (39 agents), and (ii) the other around three epistemic communities whose intents are \emph{``neuron''} (70 agents), \emph{``brain''} (36 agents) and \emph{\{``nervous'', ``system''\}} (28 agents), with many common agents and which altogether makes a group of 84 single agents.
Interestingly, there are only 15 agents common to both communities (i) and (ii), so 108 agents are well divided between the two. It is not fortuitous to see that these groups correspond exactly to what the litterature describes as signi\-ficant subfields explicitly\footnote{At the beginning of the 90's, according to Grunwald \& Eisen \cite{grun:head}, \emph{``among the first mutants to be isolated was one that was later discovered to be deficient in a \emph{growth factor} needed for axis determination, a second deficient in myofibril organization, and a third in which a specific portion of its \emph{nervous system} failed to form''}.} as well as implicitly\footnote{According to the program of the first conference on zebrafish development and genetics at the CSH Laboratory in 1994, there were seven theme-based sessions, including two on nervous system and one on growth control - so, approximately, these two fields represented half the sessions and half the community.}.
 
 Some other much smaller communities help structuring further the field: the epistemic community based on \emph{\{``toxicity''\}} is made of 23 agents with 9 shared with ``growth'' and only three with ``brain'' -- this group might be related to the study of the toxic effect of growth factors.  The epistemic community based on words \emph{``acid''} (45 agents) has an interesting descent, \emph{\{``acid'', ``amino''\}} (22 agents) and \emph{\{``acid'', ``retino''\}} (21 agents), with only 3 agents in common in the extent of \emph{\{``acid'', ``amino'', ``retino''\}}, so this is a diamond with no relation between people working on/with amino acid and retinoic acid. Also, the closed couple with intent \emph{\{``spinal'', ``cord''\}} (28 agents) includes the one based on \emph{\{``spinal'', ``cord'', ``neural'', ``ventral''\}} (20 agents) with almost as many agents, suggesting that (i) ``spinal'' and ``cord'' cannot be dissociated and (ii) people working on spinal cord are also very familiar with concepts ``neural'' and ``ventral''. 
\end{enumerate}

\begin{figure}
\includegraphics[width=1.00\linewidth
]{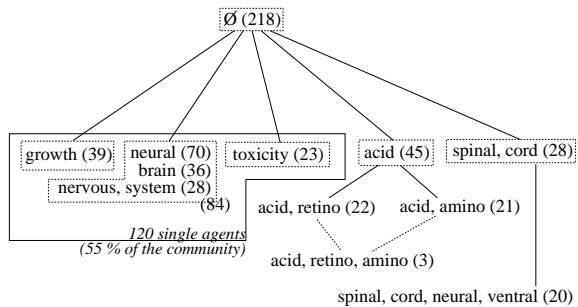}
\caption{\label{fig:partition} \small Very partial view of the actual GL (which contains more than 200,000 closed couples) hierarchically showing intents and extent sizes (in brackets) of selected epistemic communities. Note that there are various possible partitions of the whole agent set, depending on what one is looking at: for example objects, processes, methods, etc.}
\end{figure}

All these findings are summed up on figure \ref{fig:partition} and show that GLs
are efficient both for determining the community paradigm (or common background) and for finding prevailing communities as well as medium-level subcommunities. A further study would consist in observing how the community evolved through the dynamics of the GL (see section \ref{sec:dyncom}), as this embryo of partition is made from data of the period 1990-1995 and is supposed to be a \emph{static} photograph of the community structure as of December 1995, certainly appreciably different now for some ``fashionable'' subfields may have been abandoned while others have appeared. 

\paragraph{Other findings and prospects}
From the random case results we can also derive that distributions of links between agents and concepts do not alone account for the special embedded clustered structure we observe -- this result is neither surprising nor new (see for instance \cite{guel:yeas}). Nevertheless, it would be interesting to see which class of random relations (or random bipartite graphs between agents and concepts) \emph{if any} can produce the same kind of GL as in our empirical case: other properties might contribute to this structure, such as e.g. assortativity, clustering coefficient etc. In other words, how does the existence of real communities actually translate in terms of properties in relation matrix $M$, apart from a given distribution of links on rows and columns, between agents and concepts ?

Moreover, these results show the usefulness of binding social and conceptual networks and taking into account data from both networks, as proposed previously in \cite{roth:bind}, since we have communities here that are not socially linked and certainly would have been uneasy to detect -- if not impossible -- with single-network based methods (namely, based on the social network): it would be interesting to compare GL-based communities with those obtained from single-network data, in particular, see whether a single-network community is included or not in a GL-community.
Finally, considering that linguistic assumptions and processing were very poor, these preliminary findings  are also very encouraging in the prospect of improving both data quality and criteria for detecting communities (see section \ref{sec:improvements} and \ref{sec:improvementscrit}).

\section{Further directions}

\subsection{Dynamic community monitoring} \label{sec:dyncom}
Having yet categorized epistemic communities on a static basis, it would be interesting to have an account of their dynamics: we describe here how particular field evolutions could translate into properties both of epistemic communities and of the GL.

\paragraph{Field progress and specialization}
We could easily monitor 
(i) the \emph{progress or decline of a field} characterized by a given concept set, by observing respectively an increase or a decrease of the corresponding agent set (i.e. a variation in the size of the population dealing with this concept set); and (ii)
the \emph{specialization or generalization of an epistemic community} and in particular its agent set, by observing respectively an increase or a decrease of its corresponding concept set (i.e. a variation on the concept set this given agent set is working on).

\paragraph{New fields}

Alternatively, one could monitor the emergence of new fields, being either entirely new fields, or fields stemming from already existing fields (namely new interdisciplinary or multidisciplinary fields). The latter is the case where diamonds emerge or grow: the epistemic communities at the top or the bottom of a diamond are increasing in agent set size. More precisely, we distinguish two  cases: \begin{enumerate}[(i)]
\item emergence of a new ``multidiscipline'': the regrouping of two existing fields under a more general epistemic community containing agents from the two former fields. This happens when the epistemic community based on the union of two agent sets $S_{1}\cup S_{2}$ is growing, thus having $S_{1}^{\wedge}\cap S_{2}^{\wedge}$ as concept set -- in our exemple fig. \ref{fig:diamond}, it would correspond to the growth of the ``cognitive science'' community (diamond's top).
\item emergence of a new ``interdiscipline'': merging of two existing fields in a more specific epistemic community with concepts from the two former fields (growth of the epistemic community based on the union of two concept sets $C_{1}\cup C_{2}$, with $C_{1}^{\star}\cap C_{2}^{\star}$ as agent set -- e.g. the ``cognitive neurolinguistics'' community on fig. \ref{fig:diamond}, i.e. diamond's bottom).
\end{enumerate}


\subsection{Linguistic processing}\label{sec:improvements}

The improvement of linguistic processing is most urgent, and could first include the use of:
\begin{itemize}
\item Lemmatizers: algorithms giving the root of a word, instead of using a stemmer like the one used here (the ``Porter stemmer'', though it is also a quite simple yet efficient lemmatizer);
\item Taggers: algorithms detecting word grammatical status in context, e.g. ``subject'', ``verb'', etc.;
\item Morphological analyzers: algorithms recognizing the shape of a word actually composed of two or more words, like "molecular biology", "positon emission tomography", etc.;
\item Dictionaries: ontologies of the domain, returning classes of words considered as equivalent (as stated in section \ref{sec:empirical}), like ``zebrafish'' and ``rerio brachydanio'', the former being the common name of the latter;
\item Disambiguators: algorithms determining the meaning of words by examining the context in which they are used \cite{wang:usin}. 
\end{itemize}
Most of these tools already exist, although their joint use would require a judicious work of integration.


\paragraph{Expert-processed data} Alternatively, it could be useful to compare these results with those from data processed by human experts, where all linguistic processing problems become quite obsolete. For instance, (i) by providing them with a fixed list of concepts and making them classify agents according to this list, or (ii) by making them identify a restricted list of words they know to be sufficiently descriptive for a given set of articles (e.g. protein nomenclature consisting of very specific names \cite{lelu:extr}).

\subsection{Community detection criteria}\label{sec:improvementscrit}

The design of better criteria in order to categorize and distinguish \emph{medium-level} epistemic communities is also a critical question. In this paper, we used the agent set size, which is actually a quite simple criterion bearing some major drawbacks, such as the fact that small communities are ignored, even if they correspond to well-defined though isolated fields. In this respect, taking the communities which are close to the top (also called \emph{anti-chains}) can prove more relevant for they are just more specific than the whole community, obviously the most general epistemic community. In a more general view, before designing efficient criteria, it is most important to find the properties that make an epistemic community be a ``medium-level'' community; obviously the property of gathering an important proportion of the agents is a good yet insufficient first estimate. Hence, a more detailed set of properties might for instance include (i) distance from the top epistemic community, (ii) distance from the empty epistemic community ($\emptyset$,C), and (iii) concept set size.


\paragraph{GL handling} In the prospect of making this method available to scientists, a complementary approach could be to design a software allowing navigation through the lattice, like for instance starting from the top community and progressively narrowing the agent set by specifying concepts from a list of possible choices.

\section*{Conclusion}

In this paper we proposed a method for describing and categorizing communities of knowledge as well as capturing essential stylized facts regarding their structure. Assuming that such communities are structured in fields and subfields of common concerns, we aimed eventually at rebuilding this structure and in particular at providing an accurate taxonomy by automatically partitioning the community into various hierarchic representative subfields.

After having reviewed some definitions of knowledge communities or ``epistemic communities'' from social epistemology and economics, we introduced yet a definition that reflected the exact property of belonging to the same community when  sharing the same concerns and working on the same concepts --- a conception close to  structural equivalence. For a GL contains exactly all such epistemic communities, we showed next that the Galois lattice structure was a particularly adequate clustering method with respect to this definition. However, it was unclear whether this was sufficient to make it an useful categorizing tool in that the set of all epistemic communities could possibly prove really huge and intractable. To this end, we conjectured that if knowledge fields did indeed exist there should be a gap in agent set size between epistemic communities corresponding to real subfields and others (the former gathering many more agents); this first criterium will then have allowed us to discriminate within the lattice between ``uninteresting'' communities and significant ones. The lattice was thus expected to provide the hierarchic structure we wanted to rebuild.

Empirical results on an embryologist community centered around the model animal zebrafish confirmed this expectation even though data quality was somewhat imperfect, mostly because of an approximative linguistic processing. High-size epistemic communities were significantly numerous, especially with respect to selected random cases, and we managed to reproduce a partition of the community (figure \ref{fig:partition}) confirmed relevant by domain experts.

Our method diverges essentially from single-network-based methods using for instance relationships or semantic proximity, for it lies on the very duality of epistemic communities (agents having common interests) -- it would nevertheless be interestingly compared to results obtained through these other clustering methods. 
Also, it could also be fruitfully applied in other contexts such as the field of technological cooperation between companies through contracts, equivalent to authors working on concepts through articles. Several improvements could be carried out, such as better linguistic processing, better criteria design, and better handling of the lattice. Finally, as we endeavored to define, describe and hierarchize epistemic communities, a further work will attempt to explain how we could monitor their dynamics and the coevolution of the social and conceptual structures.

\paragraph{Acknowledgements}
\small

The authors wish to thank Nadine Peyri\'eras for very fruitful discussions and comments. 


\bibliography{../cambiblio}

\begin{thebibliography}{10}

\bibitem{barb:ordr}
M.~Barbut and B.~Monjardet.
\newblock {\em Alg\`ebre et Combinatoire}, volume~II.
\newblock Hachette, Paris, 1970.

\bibitem{bata:gene}
V.~Batagelj, A.~Ferligoj, and P.~Doreian.
\newblock Generalized blockmodeling.
\newblock {\em Informatica}, 23:501--506, 1999.

\bibitem{birk:latt}
G.~Birkhoff.
\newblock {\em Lattice Theory}.
\newblock American Mathematical Society, Providence, RI, 1948.

\bibitem{cowa:expl}
R.~Cowan, P.~A. David, and D.~Foray.
\newblock The explicit economics of knowledge codification and tacitness.
\newblock {\em Industrial \& Corporate Change}, 9(2):212--253, 2000.

\bibitem{dore:part}
P.~Doreian and A.~Mrvar.
\newblock A partitioning approach to structural balance.
\newblock {\em Social Networks}, 18(2):149--168, 1996.

\bibitem{dupo:orga}
O.~Dupouet, P.~Cohendet, and F.~Creplet.
\newblock {\em Economics with Heterogenous Agents}, chapter Organisational
  innovation, communities of practice and epistemic communities: the case of
  Linux.
\newblock Springer, Berlin, 2001.

\bibitem{free:usin}
L.~C. Freeman and D.~R. White.
\newblock Using galois lattices to represent network data.
\newblock {\em Sociological Methodology}, 23:127--146, 1993.

\bibitem{frie:theo}
N.~E. Friedkin.
\newblock Theoretical foundations for centrality measures.
\newblock {\em American Journal of Sociology}, 96(6):1478--1504, 1991.

\bibitem{gant:algo}
B.~Ganter.
\newblock Two basic algorithms in concept analysis.
\newblock Technical Report preprint \#831, TH-Darmstadt, 1984.

\bibitem{gier:scie}
R.~Giere.
\newblock Scientific cognition as distributed cognition.
\newblock In C.~et~al., editor, {\em The Cognitive Basis of Science}, pages
  285--299. Cambridge University Press, 2002.

\bibitem{godi:meth}
R.~Godin, G.~Mineau, R.~Missaoui, and H.~Mili.
\newblock Méthodes de classification conceptuelle basées sur les treillis de
  galois et applications.
\newblock {\em Revue d'Intelligence Artificielle}, 9(2):105--137, 1995.

\bibitem{grun:head}
D.~J. Grunwald and J.~S. Eisen.
\newblock Headwaters of the zebrafish -- emergence of a new model vertebrate.
\newblock {\em Nature Rev. Genetics}, 3(9):717--724, 2002.

\bibitem{guel:yeas}
N.~Guelzim, S.~Bottani, P.~Bourgine, and F.~K\'ep\`es.
\newblock Topological and causal structure of the yeast transcriptional
  regulatory network.
\newblock {\em Nature Genetics}, 31(5):60--63, 2002.

\bibitem{haas:intr}
P.~Haas.
\newblock Introduction: epistemic communities and international policy
  coordination.
\newblock {\em International Organization}, 46(1):1--35, winter 1992.

\bibitem{hart:kmea}
J.~A. Hartigan.
\newblock {\em Clustering Algorithms}.
\newblock Wiley, New York, 1975.

\bibitem{hopc:natu}
J.~E. Hopcroft, O.~Khan, B.~Kulis, and B.~Selman.
\newblock Natural communities in large linked networks.
\newblock In {\em Proceedings of the 9th SIGKDD International Conference on
  Knowledge Discovery and Data Mining}, pages 541--546. ACM, 2003.

\bibitem{jack:foun}
R.~Jackendoff.
\newblock {\em Foundations of Language: Brain, Meaning, Grammar, Evolution}.
\newblock Oxford University Press, 2002.

\bibitem{jacq:mode}
C.~Jacquelinet, O.~Bodenreider, and A.~Burgun.
\newblock Modelling syllepse in medical knowledge bases with application in the
  domain of organ failure and transplantation.
\newblock In {\em Proceedings of OntoLex, Workshop on Ontologies and Lexical
  Knowledge Bases 2000}, 2000.

\bibitem{john:hier}
S.~C. Johnson.
\newblock Hierarchical clustering schemes.
\newblock {\em Psychometrika}, 2:241--254, 1967.

\bibitem{klei:inte}
J.~T. Klein.
\newblock {\em Interdisciplinarity: History, Theory, and Practice}.
\newblock Wayne State University Press, Detroit, 1990.

\bibitem{koho:soms}
T.~Kohonen.
\newblock {\em Self-Organizing Maps}.
\newblock Springer, Berlin, 3rd edition, 2000.

\bibitem{kuhn:stru}
T.~S. Kuhn.
\newblock {\em The Structure of Scientific Revolutions}, chapter The Priority
  of Paradigms.
\newblock University of Chicago Press, 2nd edition, 1970(1962).

\bibitem{lavi:syst}
R.-J. Lavie.
\newblock Systemic productivity must complement structural productivity.
\newblock In {\em Proceedings of the Xth Congress of Cognitive Linguistics},
  2003.

\bibitem{lelu:extr}
A.~Lelu, P.~Bessi\`eres, A.~Zasadzinski, and D.~Besagni.
\newblock Extraction de processus fonctionnels en g\'en\'etique des microbes
  \`a partir de r\'esum\'es medline.
\newblock In {\em Proceedings of the Journ\'ees francophones d'Extraction et de
  Gestion des Connaissances (EGC)}, 2004.

\bibitem{lind:conc}
C.~Lindig.
\newblock Concepts, a free and portable implementation of concept analysis in
  c.
\newblock
  \small\mbox{www.st.cs.uni-sb.de/$\sim$lindig/src/concepts-0.3f.tar.gz}, 1998.

\bibitem{lorr:stru}
F.~Lorrain and H.~C. White.
\newblock Structural equivalence of individuals in social networks.
\newblock {\em Journal of Mathematical Sociology}, 1(49--80), 1971.

\bibitem{moll:rand}
M.~Molloy and B.~Reed.
\newblock A critical point for random graphs with a given degree sequence.
\newblock {\em Random Structures and Algorithms}, 161(6):161--179, 1995.

\bibitem{mood:stru}
J.~Moody and D.~R. White.
\newblock Structural cohesion and embeddedness: a hierarchical conception of
  social groups.
\newblock {\em American Sociological Review}, 68(103--127), 2003.

\bibitem{newm:dete}
M.~E.~J. Newman.
\newblock Detecting community structure in networks.
\newblock {\em European Phys. Journal B}, 38:321--330, 2004.

\bibitem{port:algo}
M.~F. Porter.
\newblock An algorithm for suffix stripping.
\newblock {\em Program}, 14(3):130--137, 1980.

\bibitem{radi:defi}
F.~Radicchi, C.~Castellano, F.~Cecconi, V.~Loreto, and D.~Parisi.
\newblock Defining and identifying communities in networks.
\newblock {\em PNAS}, 101(9):2658--2663, 2004.

\bibitem{roch:semi}
L.~M. Rocha.
\newblock Semi-metric behavior in document networks and its application to
  recommandation systems.
\newblock In V.~Loia, editor, {\em Soft Computing Agents: A New Perspective for
  Dynamic Information Systems}, International Series Frontiers in Artificial
  Intelligence and Applications, pages 137--163. IOS Press, 2002.

\bibitem{rosc:cogn}
E.~Rosch and B.~Lloyd.
\newblock Cognition and categorization.
\newblock {\em American Psychologist}, 44(12):1468--1481, 1978.

\bibitem{roth:bind}
C.~Roth and P.~Bourgine.
\newblock Binding social and cultural networks: a model.
\newblock arXiv:nlin.AO/0309035, 2003.

\bibitem{schm:soci}
F.~Schmitt, editor.
\newblock {\em Socializing Epistemology : The Social Dimensions of Knowledge}.
\newblock Rowman \& Littlefield, 1995.

\bibitem{spec:prob}
D.~F. Specht.
\newblock Probabilistic neural networks for classification, mapping, or
  associative memory.
\newblock In {\em Procedings of IEEE International Conference On Neural
  Networks}, pages 525--532, 1988.

\bibitem{wang:usin}
L.~Wang, W.~Song, and D.~Cheung.
\newblock Using contextual semantics to automate the web document search and
  analysis.
\newblock In {\em Proceedings of the First International Conference on Web
  Information Systems Engineering (WISE)}, 2000.

\bibitem{will:con2}
R.~Wille.
\newblock Conceptual graphs and formal concept analysis.
\newblock In {\em Proceedings of the fourth International Conference on
  Conceptual Structures}, number \#1257 in Lecture Notes on Computer Science,
  pages 290--303. Springer-Verlag, Berlin, 1997.
\newblock http://citeseer.nj.nec.com/wille97conceptual.html.

\end{thebibliography}
\bibliographystyle{abbrv}
\end{document}